\documentclass[pra,twocolumn,superscriptaddress,longbibliography]{revtex4-1}

\usepackage{amsmath,amssymb,graphicx,color,hyperref}
\hypersetup{
    colorlinks=true,
    linkcolor=blue,
    citecolor=red,
    urlcolor=blue
}
\graphicspath{{figures/}}

\begin{document}

\title{Plasmon-assisted photoelectron emission in a model cluster using  time-dependent density functional theory and the time-dependent surface-flux method}

\author{Mikhail Bednov}
\email{mikhail.bednov@uni-rostock.de}
\affiliation{Institute of Physics, University of Rostock, 18059 Rostock, Germany}

\author{Waqas Pervez}
\email{waqas.pervez@uni-rostock.de}
\affiliation{Institute of Physics, University of Rostock, 18059 Rostock, Germany}

\author{Ingo Barke}
\email{ingo.barke@uni-rostock.de}
\affiliation{Institute of Physics, University of Rostock, 18059 Rostock, Germany}
\affiliation{Department Life, Light \& Matter, University of Rostock, 18051 Rostock, Germany}

\author{Dieter Bauer}
\email{dieter.bauer@uni-rostock.de}
\affiliation{Institute of Physics, University of Rostock, 18059 Rostock, Germany}
\affiliation{Department Life, Light \& Matter, University of Rostock, 18051 Rostock, Germany}

\date{\today}

\begin{abstract}
We investigate plasmon-assisted photoelectron emission using a one-dimensional time-dependent density-functional theory (TDDFT) model. The plasmons are excited nonlinearly by three laser photons.  Photoelectron spectra are computed with the time-dependent surface-flux (t-SURFF) method. In addition to the expected above-threshold ionization (ATI) comb, we observe peaks that arise from long-lived plasmon oscillations and the associated electron emission occurring after the laser pulse. We further analyze the positions of these peaks and their scaling behavior with the laser intensity. \end{abstract}
\maketitle

\section{Introduction}

Plasmonic nanostructures concentrate optical energy into deep-subwavelength volumes, enabling strong-field photoemission, ultrafast electron sources, and nanoscale control of light-matter interactions \cite{Li2013,Putnam2017,Stockman11,Dombi2020,Kim2023}. Ultrafast photoemission electron microscopy (PEEM) has directly imaged plasmon dynamics in space and time, providing unique insight into how localized fields drive electron emission at the nanoscale \cite{Lemke_etal2013,Dabrowski,Kosar24,Gosh24}. On the theoretical side, real-time time-dependent density functional theory (TDDFT) \cite{ullrich2011} has emerged as a powerful framework for following collective electronic motion, charge transfer, and field-driven dynamics in real time and real space, complementing frequency-domain linear-response TDDFT and classical electrodynamics approaches \cite{Uchida2017,Ma2015,Herring,Wang2024}.

A plasmon is a quantized collective oscillation of electrons. In the idealized case of solving the laser-driven many-body time-dependent Schrödinger equation (TDSE) exactly, a plasmon emerges naturally as a superposition of many-body states. If some of these states correspond to autoionizing resonances, the plasmon oscillation can be accompanied by electron emission. Because the TDSE is linear, dipole oscillations occur only through such superpositions of eigenstates (including continua associated with ionization). After the laser pulse, populations remaining in autoionizing states continue to decay; however, this post-pulse decay may be negligible if the resonance lifetimes are much shorter than the pulse duration.

In TDDFT, the interacting electron system and linear TDSE are replaced by a formally exact non-interacting Kohn-Sham (KS) system described by nonlinear time-dependent Kohn-Sham (TDKS) equations. Within this framework, a plasmon corresponds to a collective oscillation of KS particles that all experience the same time-dependent KS potential. The exact KS potential, if known, would reproduce the same dipole expectation value as the full many-body solution. In practice, however, commonly used approximate functionals, such as the adiabatic local-density approximation (ALDA), tend to produce persistent oscillations after the laser pulse, as we will demonstrate in this work. Furthermore, the exact observable for photoelectron emission in TDDFT is unknown \cite{Veniard2003,Brics2013}, and it is customary to interpret the KS particles themselves as electrons when computing photoelectron spectra.

Photoelectron spectra (PES) in strong-field physics are commonly obtained using (i) projections onto field-free continuum states, (ii) the window-operator method \cite{Schafer90,Bauer2017}, or (iii) the time-dependent surface-flux method (t-SURFF) \cite{Tao2012,Bauer2017,TulskyQprop2020}. Methods (i) and (ii) are difficult to apply to TDDFT because they require a stationary potential after the laser pulse, whereas the KS potential continues to evolve. Method (iii), t-SURFF, avoids this assumption but requires a sufficiently long post-pulse propagation so that electrons of the lowest kinetic energy of interest have reached the flux surface. For the TDSE, whose Hamiltonian is stationary after the pulse, this post-propagation can be performed in a single numerical step. In TDDFT, however, the KS Hamiltonian remains time-dependent after the pulse, precluding such a  shortcut.

In this work, we show that the t-SURFF photoelectron spectra obtained within TDDFT exhibit sharp peaks whose positions and widths depend  on the length of the post-pulse propagation. These peaks originate from plasmon-assisted electron emission occurring after the pulse: the longer the post-pulse propagation, the narrower the peaks become. Although this sensitivity is clearly an artifact of approximate exchange-correlation functionals, plasmon-assisted electron emission is a genuine physical effect. This enables us to analyze, for example, the scaling behavior of plasmon-assisted electron emission, while bearing in mind that its magnitude is likely overestimated by TDDFT with simple exchange-correlation potentials.

Atomic units are used unless noted otherwise.

\section{Model}
\label{sec:methods}

We model a one-dimensional (1D) cluster of $N_e=40$ electrons confined by $N_\text{ion} = N_e $ equidistantly spaced ions, \begin{equation} V_\text{ion}(x) = - \sum_{i=1}^{N_\text{ion}} \frac{1}{\sqrt{(x-X_i)^2 + 1}} 
\end{equation} 
with $X_i = (-(N_\text{ion}+1)/2  + i) a$ and $a=1.125$. The value for $a$ was chosen such that the KS energy for the valence orbital is close to the work function of silver clusters. 

Density functional theory (DFT) is used to calculate the ground state configuration  for  $N_\text{KS}=N_e/2$ spin-degenerate KS orbitals according to the KS equation
 \begin{equation}
 \varepsilon_i  \varphi_i(x)
 =
 \Bigl[
   -\frac{1}{2}\frac{\partial^2}{\partial x^2}
   + V_\text{KS}[n](x)
 \Bigr]\,
 \varphi_i(x)
 \label{eq:TIKS}
\end{equation} 
with
\begin{equation}
V_\text{KS}[n](x) = V_\text{ion}(x) + V_\text{H}[n](x) + V_\text{xc}[n](x) \label{eq:timeindepVKS}
\end{equation}
where  \begin{equation}
V_\text{H}[n](x) = \int \text{d} x'\, \frac{n(x')}{\sqrt{(x-x')^2 + 1}}
\end{equation}
is the Hartree potential and \begin{equation}
V_\text{xc}[n](x) \simeq  V_\text{x}[n](x) = -  \left(\frac{3}{\pi} n(x)\right)^{1/3} \label{eq:xc}
\end{equation}
is the exchange-correlation  potential in exchange-only adiabatic local density approximation. Note that we do not aim at modelling a 1D electron system (such as in a nano wire, for instance) but rather a 3D electron system along the laser polarization direction. This is the reason why we use the expression \eqref{eq:xc} for a 3D electron gas in our 1D model.

\begin{figure}[t]
  \centering
  \includegraphics[width=0.9\linewidth]{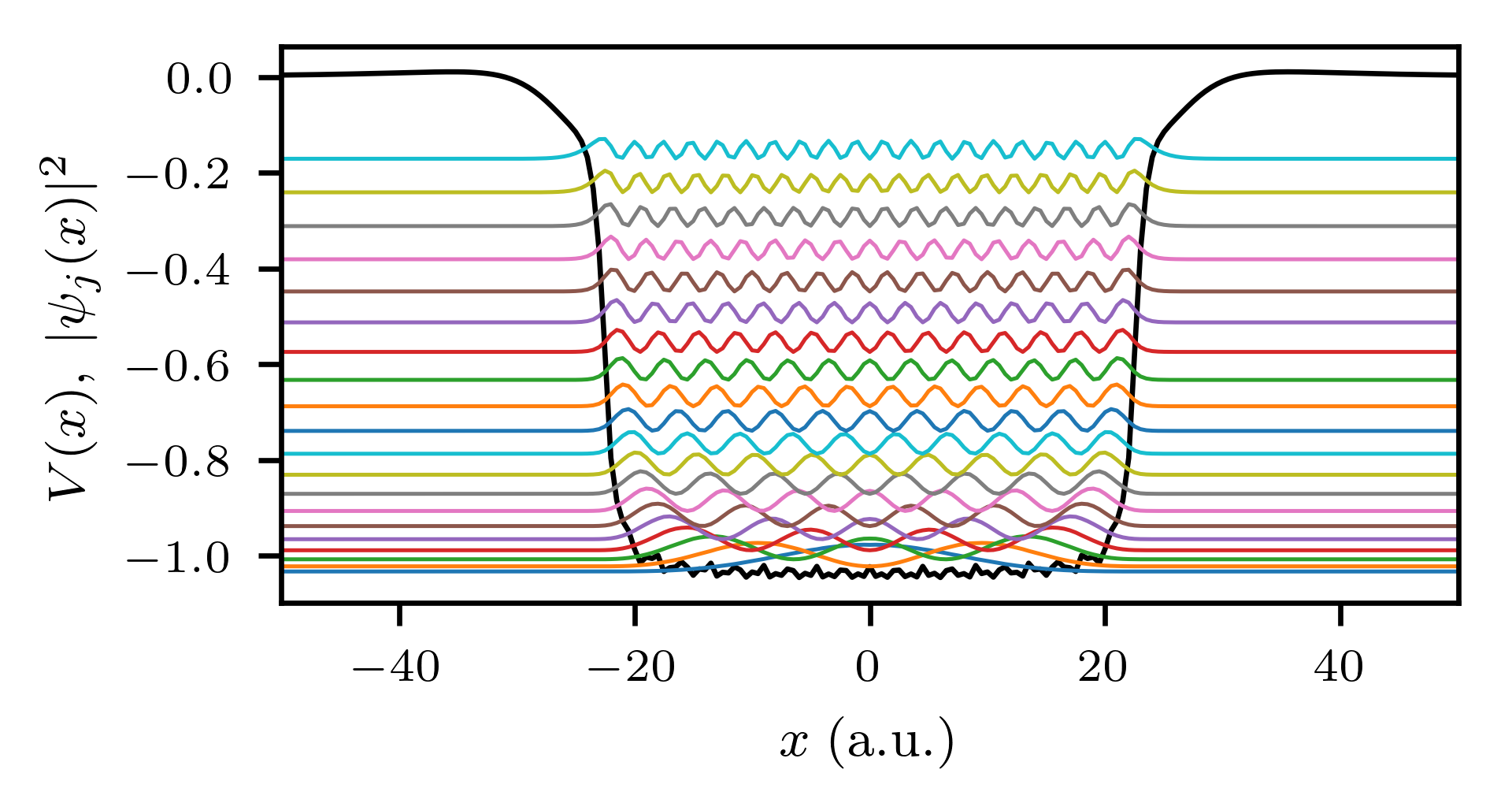}
  \caption{
    Kohn-Sham potential $V_{\rm KS}(x)$ (solid black) of the 1D  cluster and the
    20 occupied KS orbitals (coloured) plotted as $|\varphi_i(x)|^2+\varepsilon_i$.
     }
  \label{fig:VKS_levels}
\end{figure}

  We use TDDFT to simulate this system in a laser field with vector potential $A(t)$ in dipole approximation. The TDKS equation  reads
\begin{equation}
 i\partial_t \varphi_i(x,t)
 =
 \Bigl[
   \tfrac{1}{2}\bigl(-i\partial_x + A(t)\bigr)^2
   + V_\text{KS}[n](x,t)
 \Bigr]\,
 \varphi_i(x,t)
 \label{eq:TDKS}
\end{equation}
where $V_\text{KS}[n](x,t)$ corresponds to \eqref{eq:timeindepVKS} with the time-dependent density
\begin{equation}
n(x,t) = 2\sum_{i=1}^{N_\text{KS}} |\varphi_i(x,t)|^2
\end{equation}
inserted. 
 Figure~\ref{fig:VKS_levels} shows the ground state KS potential and the $20$ ground state KS probability densities $|\varphi_i(x)|^2$, shifted to their respective KS energies $\varepsilon_i$.

We apply a laser pulse with the vector potential
\begin{equation}
A(t) =
\begin{cases}
a_0 \cos(\omega_L t)
\,\sin^2\!\Bigl(\dfrac{\omega_L t}{2 N_\text{cyc}}\Bigr), & 0<t<T_p,\\[6pt]
0, & \text{otherwise},
\end{cases}
\label{eq:laser}
\end{equation}
\begin{equation}
T_p=\dfrac{2\pi N_\text{cyc}}{\omega_L}.
\label{eq:T_p}
\end{equation}
The electric field is then given by $E_L(t)=-\dot A(t)$.

We employ the t-SURFF method \cite{Tao2012,Bauer2017}  to compute photoelectron spectra. In 1D and velocity gauge, the momentum-space amplitudes collected at the left (right) surfaces $x_{L}$   ($x_{R}$) read (for each occupied orbital $\varphi_i$) 
\begin{align}
b_{R,i}(k) &= \nonumber\\ 
&\quad \frac{1}{\sqrt{2\pi}}
\int_0^\tau \text{d}t\,
e^{\frac{i}{2}k^2 t}\, e^{-ik(x_R-\,\alpha(t))}\, \nonumber\\
&\quad\Bigl[
\Bigl(\tfrac{k}{2}+A(t)\Bigr)\,\varphi_i(x_R,t)
 - \frac{i}{2}\,\partial_x\varphi_i(x_R,t)
\Bigr],
\label{eq:tsurffR}
\\
b_{L,i}(k) &= \nonumber\\
&\quad -\frac{1}{\sqrt{2\pi}}
\int_0^\tau\text{d}t\,
e^{\frac{i}{2}k^2 t}\, e^{-ik(x_L-\,\alpha(t))}\, \nonumber\\
&\quad\Bigl[
\Bigl(\tfrac{k}{2}+A(t)\Bigr)\,\varphi_i(x_L,t)
 - \frac{i}{2}\,\partial_x\varphi_i(x_L,t)
\Bigr],
\label{eq:tsurffL}
\end{align}
where $\alpha(t)=\int_0^t A(t')\,\text{d}t'$ is the classical displacement and $\tau$ is the end time of the simulation. Numerically, the time integral is performed using the trapezoidal rule with the same $\Delta t$ that is used for propagation.     The total energy spectrum is 
\begin{equation}
Y(k) = \sum_{i=1}^{N_\text{KS}} \bigl| b_{R,i}(k)+b_{L,i}(k) \bigr|^2,
\qquad
E=\frac{k^2}{2}. \label{eq:totaltsurffspec}
\end{equation}
While this way of calculating electron spectra is correct for single-particle TDSEs it is only an approximation in the TDKS context.
As mentioned in the Introduction, the density functional for the observable ``electron spectrum'' is unknown \cite{Veniard2003,Brics2013}. Treating KS orbitals as if they were single-electron wavefunctions is an approximation.

\subsection{Numerical details}
We use $N=2000$ grid points with a spacing $\Delta x=0.5$, a time step $\Delta t=0.25$, and complex absorbing boundaries at the edges. The convolution theorem is used to calculate $V_\text{H}[n](x,t)$.  Real‑time propagation of (\ref{eq:TDKS}) uses  the  Crank-Nicolson scheme with a predictor-corrector step and  absorbing boundaries \cite{Bauer2017}. The initial configuration, solving \eqref{eq:TIKS} is obtained with imaginary-time propagation using also the Crank-Nicolson scheme.
 The t-SURFF surfaces are placed at $x_L=0.25\,N\Delta x$ and $x_R=0.75\,N\Delta x$. Convergence was verified with respect to box size, time step, position of the t-SURFF surfaces, and the $k$-grid used for the electron spectra.


\section{Identification of collective modes} \label{sec:linresp}

In order to identify whether a linear-response peak is due to single-particle transitions or collective in nature, it is useful to look at the response of individual orbitals to a perturbation. To that end  we apply a weak $\delta$-kick in the electric field (i.e., a $\Theta$-step in the vector potential) to the ground state configuration. Let us first look at the   Fourier-transformed  total dipole  \begin{equation}
P(\Omega) = \Omega^4\bigl|\text{FFT}[D(t)](\Omega)\bigr|^2, \quad D(t)=\int \text{d}x \,  x n(x,t).
\end{equation}
The pre-factor $\Omega^2$ arises because we are actually interested in the FFT of $\ddot D(t)$, the dipole acceleration, so that $\Omega$ is the frequency at which the dipolar kicked system radiates.

\begin{figure}[t]
  \centering
  \includegraphics[width=\columnwidth]{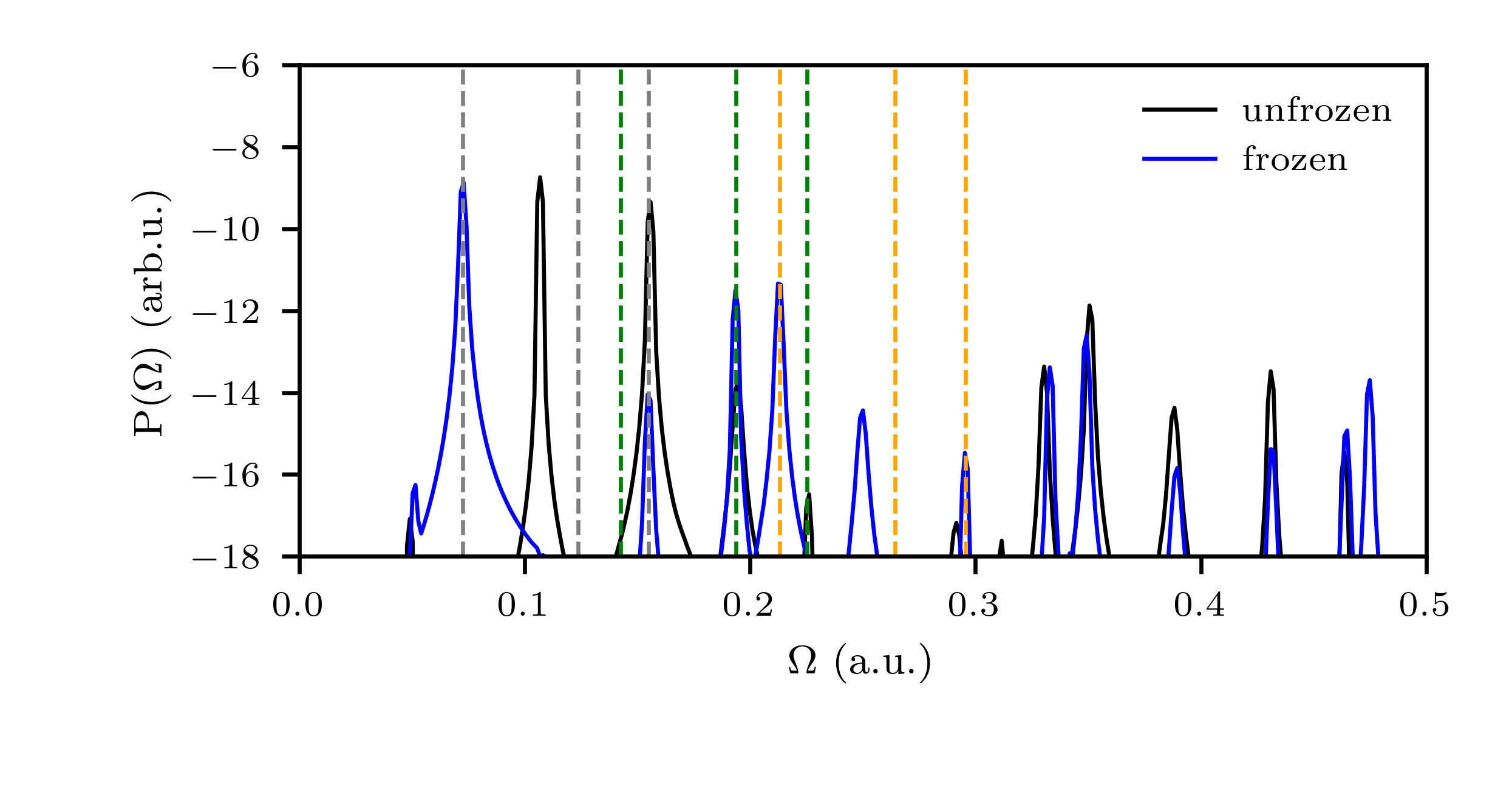}
  \caption{
  \label{fig:lin_total}
  Linear-response  spectra. Black: full, dynamic response according to  \eqref{eq:TDKS}. Blue: response for frozen KS potential $V_\text{KS}[n](x,0)$. Vertical dashed lines indicate KS level differences. Two strong peaks at $\omega_A=0.106$ and $\omega_B=0.156$ appear in the fully dynamic (``unfrozen'') response.}
\end{figure}

 Figure~\ref{fig:lin_total} shows results for the ``frozen'' and the fully dynamic system. Here, ``frozen'' means that we set $V_\text{KS}[n](x,t)$ in    \eqref{eq:TDKS} to $V_\text{KS}[n](x,0)$. In such a simulation, one expects peaks at energy differences $\varepsilon_j-\varepsilon_i$. The expected positions for such peaks are indicated by vertical dashed lines. The strongest peak is at the energy difference between the highest occupied KS orbital ($\varepsilon_{20} = -0.1709$) and the first unoccupied ($\varepsilon_{21} = -0.0984$), indicated by the first gray dashed line. There is no peak at $\varepsilon_{22}-\varepsilon_{20}$ (second gray dashed line) because of the dipole selection rule, while there is one at $\varepsilon_{23}-\varepsilon_{20}$ (third gray dashed line). The same repeats for the second-highest occupied KS orbital at $\varepsilon_{21}-\varepsilon_{19}$  (first green dashed line, no peak), $\varepsilon_{22}-\varepsilon_{19}$  (second green dashed line, peak present) etc. The yellow dashed lines indicate $\varepsilon_{21}-\varepsilon_{18}$, $\varepsilon_{22}-\varepsilon_{18}$, $\varepsilon_{23}-\varepsilon_{18}$. There are no peaks for transitions between initially populated KS levels $i,j$ because $\varepsilon_i \to \varepsilon_j$ interferes destructively with $\varepsilon_j \to \varepsilon_i$  (this is how the Pauli principle is realized in a system of non-interacting identical particles). 

\begin{figure}[t]
  \centering
  \includegraphics[width=\columnwidth]{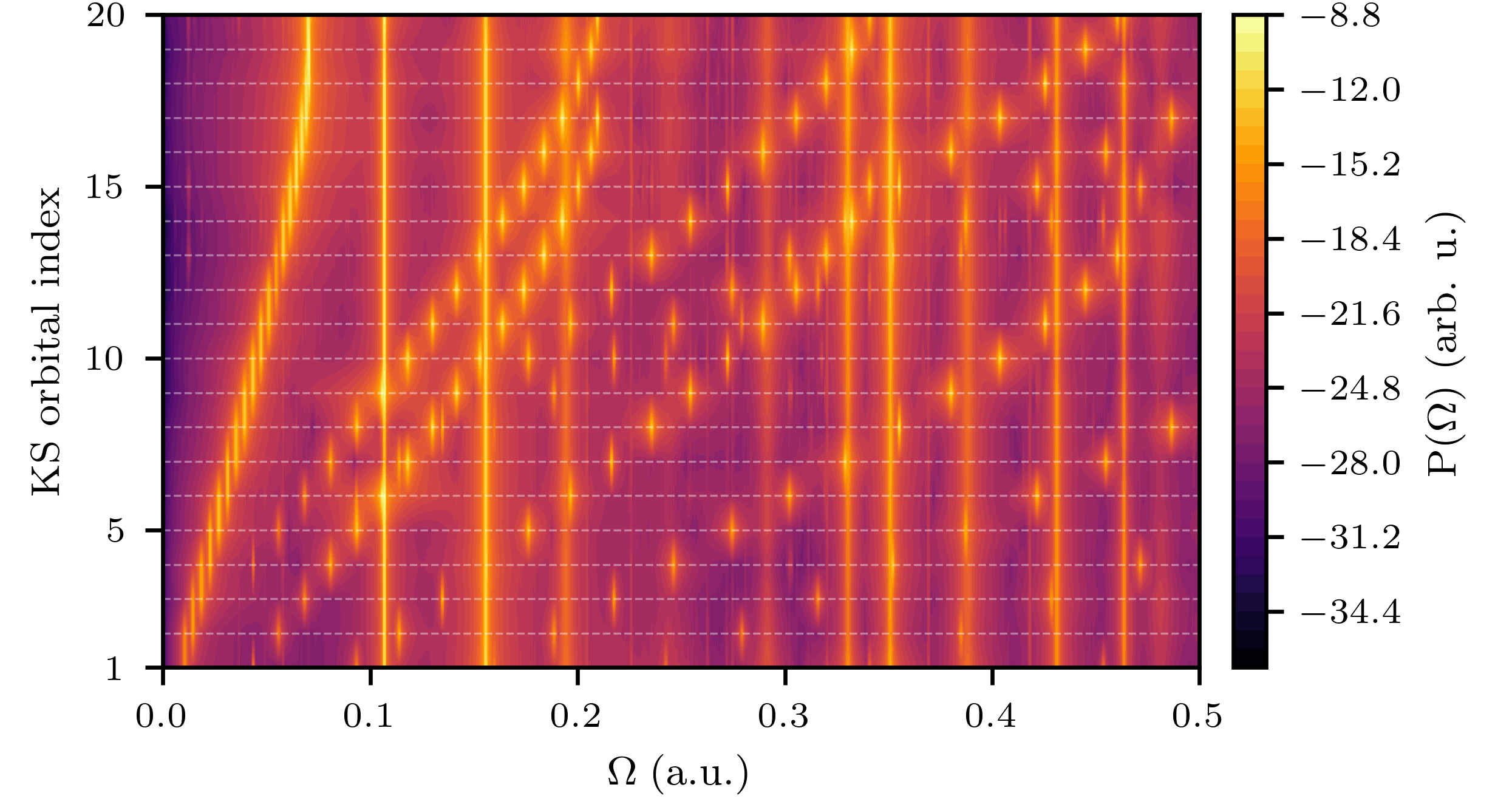}
  \caption{
  \label{fig:lin_orbital}
  Orbital-resolved spectra. A strong response of all KS orbitals at a given $\Omega$ indicates a collective mode. Instead, orbital-dependent responses are due to single-particle transitions. }
\end{figure}

In the full, dynamical response (labeled ``unfrozen'' in Fig.~\ref{fig:lin_total} )  two strong peaks at
\begin{equation}
\omega_A = 0.106, \qquad
\omega_B = 0.156
\label{eq:omegas}
\end{equation}
are observed. In order to prove that these are collective responses we show orbital-resolved linear response spectra in Fig.~\ref{fig:lin_orbital}. In such spectra, one sees all the transitions to other states, which, however, are different for the different KS orbitals. Only in the case of a harmonic-oscillator-like KS potential with equidistant levels would there  be only one common response at the harmonic oscillator frequency for all orbitals. However, we do see response peaks in  Fig.~\ref{fig:lin_orbital}  that are at the same $\Omega$ for all KS orbitals despite the fact that $V_\text{KS}[n](x)$ is not a harmonic-oscillator potential.  By definition, this is a collective response, as all KS orbitals oscillate in  unison with this frequency. The first two of such collective responses are at $\Omega=\omega_A$ and $\omega_B$.

\begin{figure}[t]
  \centering
  \includegraphics[width=\columnwidth]{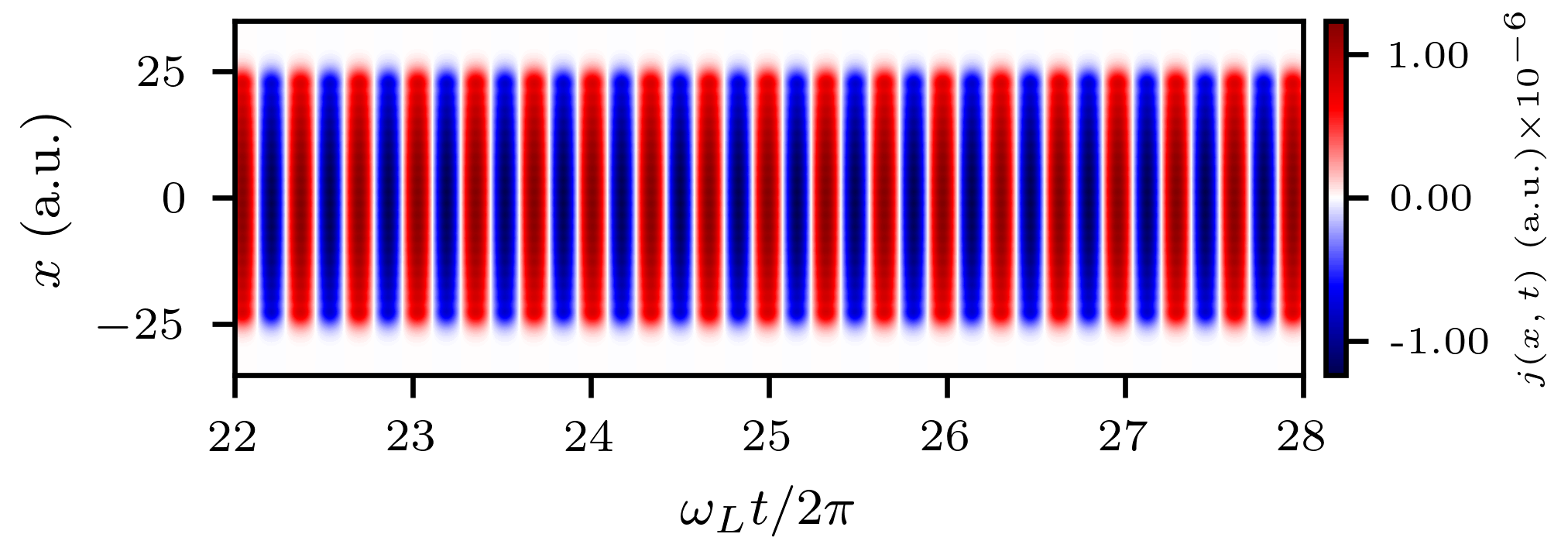}\\
  \includegraphics[width=\columnwidth]{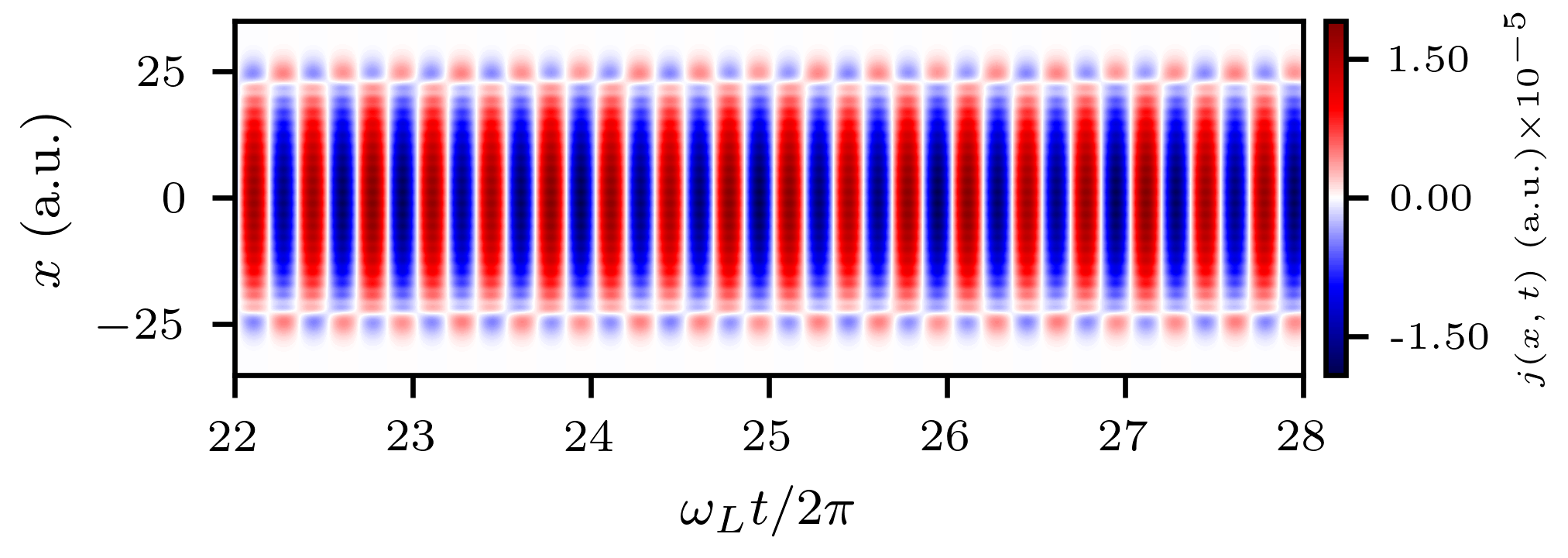}
  \caption{
  \label{fig:currents}
  Oscillating current densities for the collective modes $\omega_A = 0.106$ (upper panel) and  $\omega_B = 0.156$ (lower panel) as a function of time (in units of $2\pi/\omega_L$) and space. The modes were excited with laser pulses of frequencies $\omega_L=\omega_A/3$ and $\omega_L=\omega_B/3$, respectively. The oscillating current densities are shown for times  after the laser pulses.  }
\end{figure}

The spatial and temporal evolution of the current density for the two collective modes is shown in Fig.~\ref{fig:currents}. The oscillations are plotted for times following the laser pulses that selectively excited these modes, using laser frequencies of $\omega_L=\omega_A/3$ and $\omega_L=\omega_B/3$, respectively. Although these modes could in principle also be excited by single-photon absorption, we focus in this work on photoelectron emission following {\em nonlinear} plasmon excitation. Since $2\omega_L$ excitations are forbidden by selection rules, the lowest-order nonlinear plasmon excitations occur at $3\omega_L$. As expected,  Fig.~\ref{fig:currents} shows that the lower-energy mode $\omega_A$ corresponds to a collective oscillation of the entire electron density, while the higher-energy mode exhibits two nodes near the edges of the one-dimensional cluster. Consequently, the current density inside the cluster oscillates in the opposite direction to the surface currents.

\section{Photoelectron spectra}
\label{sec:tsurff}
\begin{figure}[t]
  \centering
  \includegraphics[width=1\linewidth]{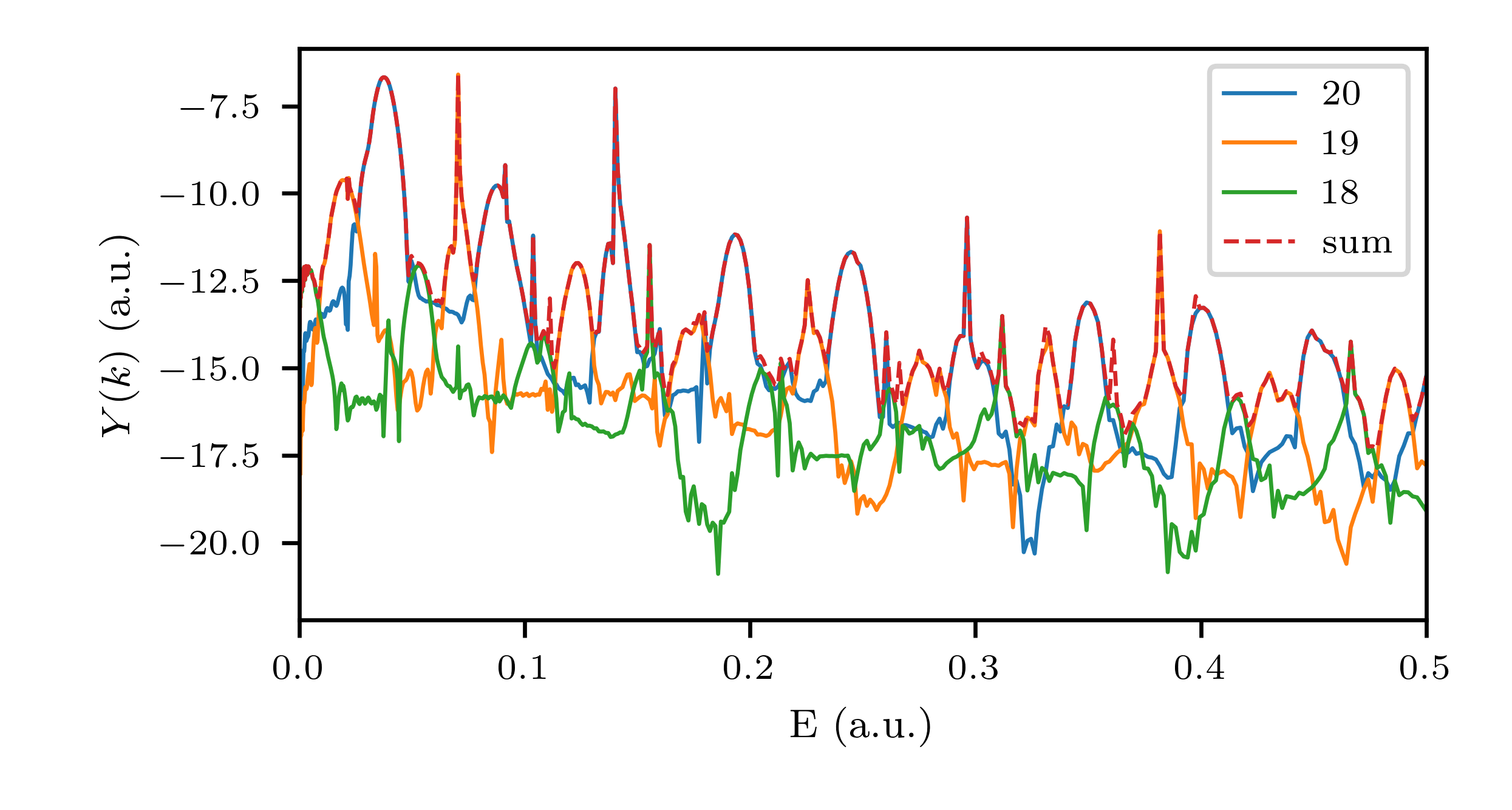}
  \caption{
    \label{fig:tsurff_total}
    Total photoelectron spectrum (red) and  the individual contributions from the three highest initially populated KS orbitals for a pulse that dominantly excites $\omega_B = 0.156$ with three laser photons.   Narrow spikes are equally spaced by $\omega_B$. Laser parameters are $a_0 = 0.004$, $\omega_L = 0.052=\omega_B/3$, $N_\text{cyc} =20$.
  }
\end{figure}

Figure~\ref{fig:tsurff_total} shows the photoelectron spectrum \eqref{eq:totaltsurffspec}   for a laser pulse with parameters $a_0 = 0.004$, $\omega_L = 0.052 = \omega_B/3$, and  $N_\text{cyc} =20$. Also shown are the individual spectra for the highest occupied KS orbitals $i=20,19,18$.  Lower-lying KS orbitals do not play a role for the total electron yield. Besides the expected ATI peaks, separated by $\omega_L$, there are very sharp spiky features. The more detailed analysis presented below shows that these features arise from plasmon excitations that decay too slowly in the TDDFT treatment. Extended post-pulse propagation within the t-SURFF framework is required to collect low-energy electrons at the distant t-SURFF surfaces. During this propagation, t-SURFF also accumulates probability amplitude associated with plasmon-assisted electron emission occurring after the laser pulse. The longer the post-pulse duration, the more pronounced and sharper these peaks become.

The laser frequency used for Fig.~\ref{fig:tsurff_total} predominantly excites the plasmon with frequency  $\omega_B$ with three laser photons of energy $\omega_L = 0.052$. In this case, the sharp peaks appear at energies $ 
E^{(B)}_{in} = \varepsilon_i + n\,\omega_B$ with $n \ge \left\lceil |\varepsilon_i|/\omega_B \right\rceil$. If the laser excites predominantly the $\omega_A$-plasmon one observes peaks at $E^{(A)}_{in} = \varepsilon_i + n\,\omega_A$.

\subsection{Time-of-flight analysis}
The t-SURFF equations \eqref{eq:tsurffR}, \eqref{eq:tsurffL} allow for a time-frequency analysis. Instead of performing the time integration from $t=0$ until the end of the simulation $\tau$ we apply a Gaussian time window of a certain width and  centered at $t$  and single-out contributions to the electron spectra that arrive within this time window at the t-SURFF surfaces. Calculating the time-of-flight for the final energy, one can trace back when the corresponding electrons were emitted. This method is similar to the Gabor transformation where the time window is applied to a Fourier transform, which is often used for the analysis of high-harmonic generation. The t-SURFF time-of-flight method was introduced in  \cite{TulskyPRA2020,Sanchez23} to investigate intra-cycle strong-field ionization.

\begin{figure}[t]
  \centering
  \includegraphics[width=\columnwidth]{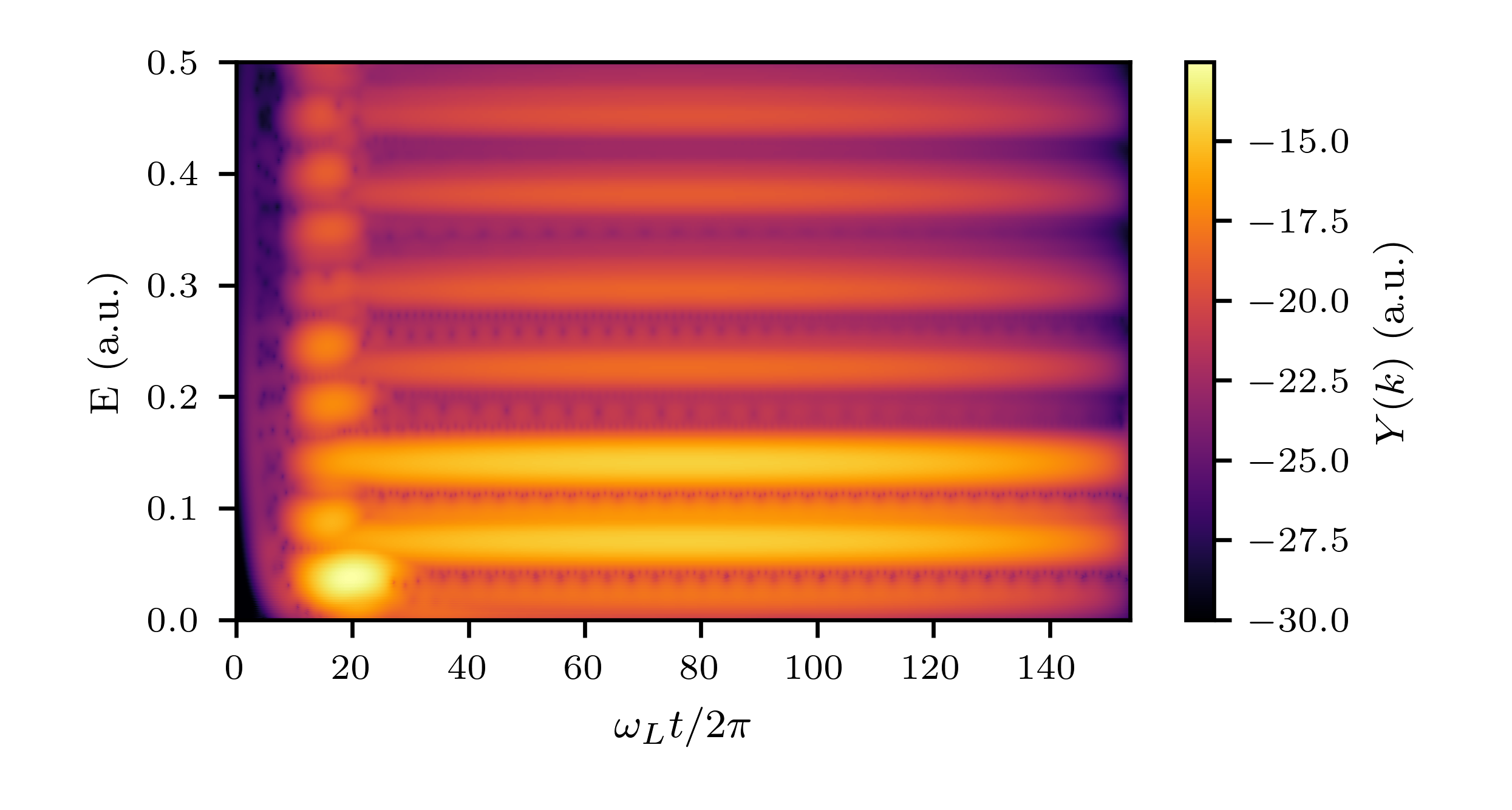}
  \caption{
    \label{fig:gabor}
     Windowed t-SURFF spectrum as a function of time (in units of $2\pi/\omega_L$) and energy $E=k^2/2$.
    Laser parameters are the same as in Fig.~\ref{fig:tsurff_total}.
  }
\end{figure}

Figure   \ref{fig:gabor} shows the windowed t-SURFF spectrum for the same laser parameters as in Fig.~\ref{fig:tsurff_total}. The width of the Gaussian time-window was one laser cycle so that the ATI peaks due to direct,  laser-induced ionization are clearly visible. All ATI-emitted electrons arrive before $\omega_L t/2\pi \simeq 30$ at the t-SURFF surfaces, which is consistent with them being emitted during the laser pulse plus  their time-of-flight. After the ATI peaks, the plasmon-assisted emission builds up and  is clearly visible as horizontal stripes at positions $ 
E^{(A,B)}_{in} = \varepsilon_i + n\,\omega_{A,B}$. The stripes are broader in energy than the narrow spikes in the total spectrum in Fig.~\ref{fig:tsurff_total} because of the narrow time-window. This also shows that the plasmon-assisted electron emission peaks would be broader (and probably closer to reality) if we stop the simulation soon after the laser pulse. However, at that time the low-energetic electrons have not yet arrived at the t-SURFF surfaces and thus are missing in the spectra. We also experimented with ``freezing'' the KS potential after the laser pulse, which, however, yields other artefacts.

\subsection{Scaling of the yield with laser intensity}
In the perturbative regime, we expect an $I^n$-scaling of the electron yield with respect to the laser intensity $I=E_L^2$ and the number $n$ of photons absorbed. In order to test this, we solve the TDKS equations for three vector potential amplitudes $a_0$, pick an ATI peak and measure the scaling of its electron yield $Y(k)$ in order to determine $n$. Then we do the same for exemplary plasmon-related peaks.

\begin{figure}[t]
  \centering
  \includegraphics[width=\linewidth]{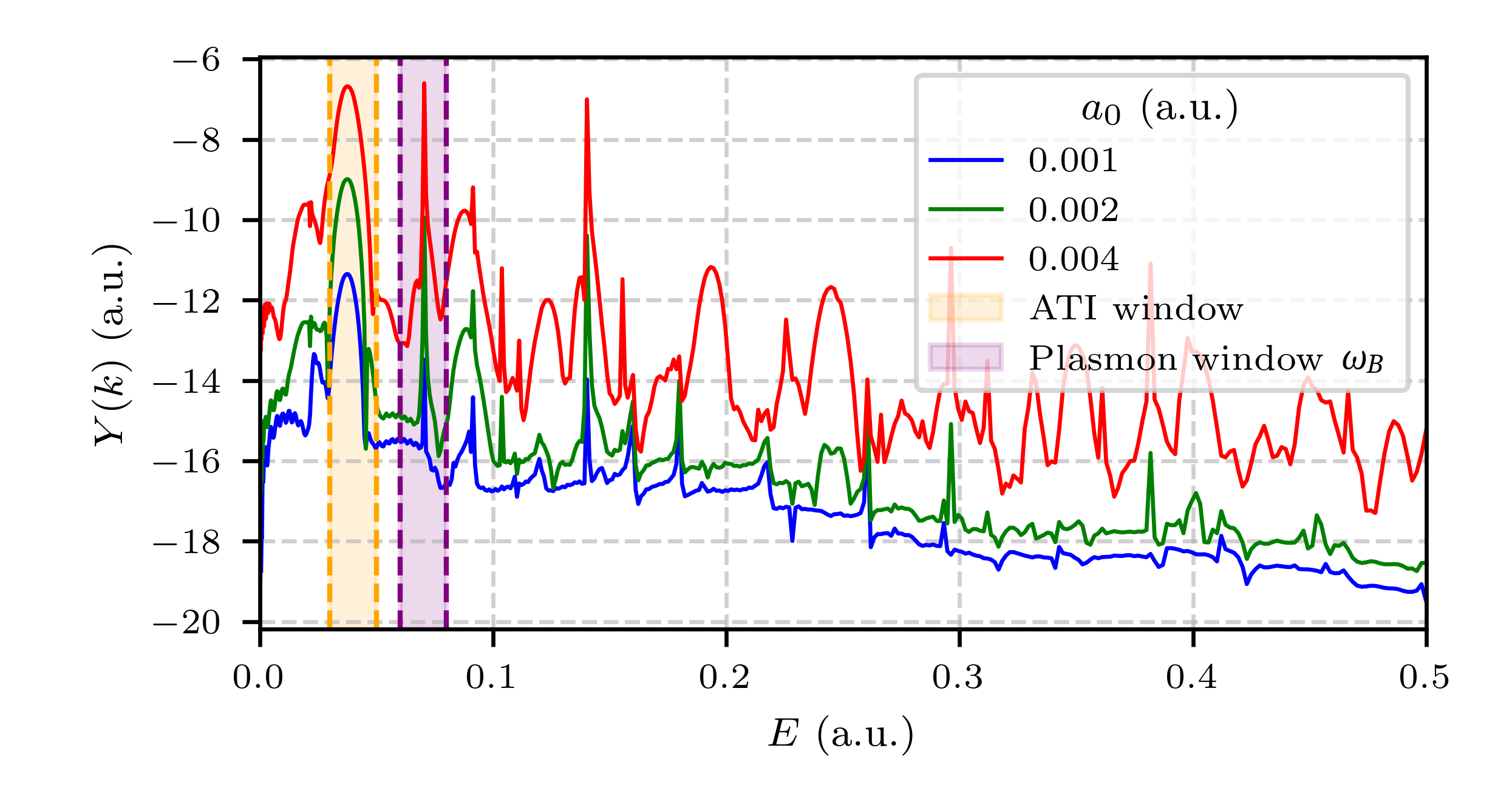}
  \caption{
    \label{fig:tsurff_comp_052}
    t-SURFF spectra for $\omega_L=0.052$, $N_\text{cyc} =20$, and three different vector potential amplitudes as given in the legend.   Shaded regions mark the ATI and plasmon windows used for determining the $I^n$-scaling.} 
\end{figure}

\begin{figure}[t]
  \centering
  \includegraphics[width=\linewidth]{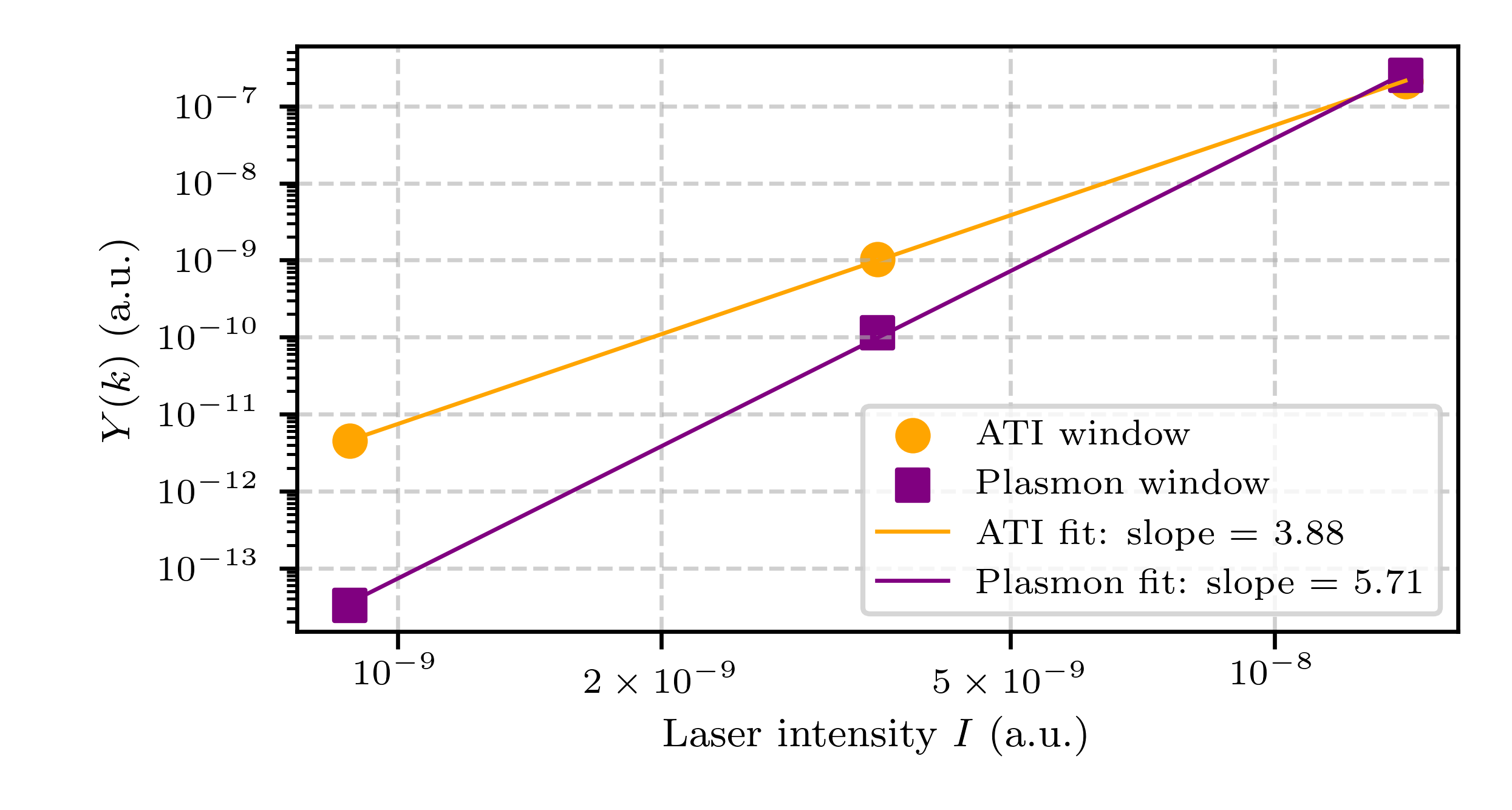}
  \caption{
    \label{fig:int_comp_052}
    Peak yields {\em vs} intensity  for Fig.~\ref{fig:tsurff_comp_052}. The ATI peak scales with $n=3.88$ while the plasmon peak scales with $n=5.71$.}
\end{figure}

Figure~\ref{fig:tsurff_comp_052} shows for  $\omega_L = 0.052=\omega_B/3$, $N_\text{cyc} =20$ photoelectron spectra for the three different $a_0$ given in the legend. We analyze the ATI peak at energy $E=0.037$, which arises from $\varepsilon_{20}+ 4\omega_L$. Hence we expect an $n=4$ scaling, which is confirmed in Fig.~\ref{fig:int_comp_052} where we find numerically $n=3.88$. We also examine the ``plasmon peak'' (meaning, ``peak due to plasmon-assisted electron emission'') at $E=0.071$, corresponding to  $\varepsilon_{19}+ 2\omega_B$. The numerical measurement gives an $n=5.71$-scaling for this peak. One could explain this by observing that it takes three laser photons to excite the $\omega_B$-plasmon. The plasmon then acts like a laser itself, and the electron takes one plasmon energy more to escape from the cluster. The two $\omega_B$ then correspond to six $\omega_L$, which is close to the numerically determined $n=5.71$.

\begin{figure}[t]
  \centering
  \includegraphics[width=\linewidth]{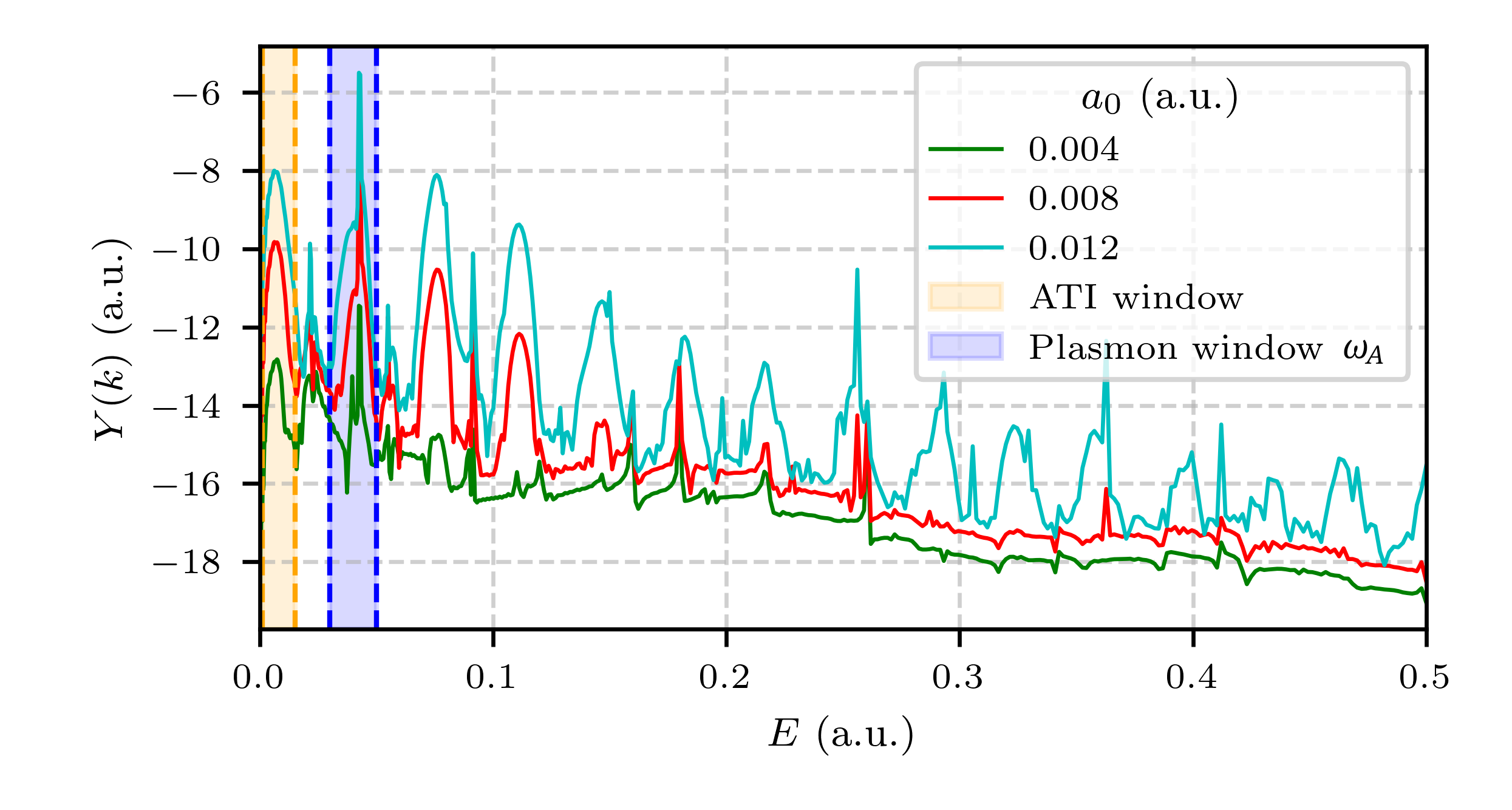}
  \caption{
    \label{fig:tsurff_comp_035} t-SURFF spectra for $\omega_L=0.035$, $N_\text{cyc} =20$, and three different vector potential amplitudes as given in the legend.   Shaded regions mark the ATI and plasmon windows used for determining the $I^n$-scaling.
    }
\end{figure}

\begin{figure}[t]
  \centering
  \includegraphics[width=\linewidth]{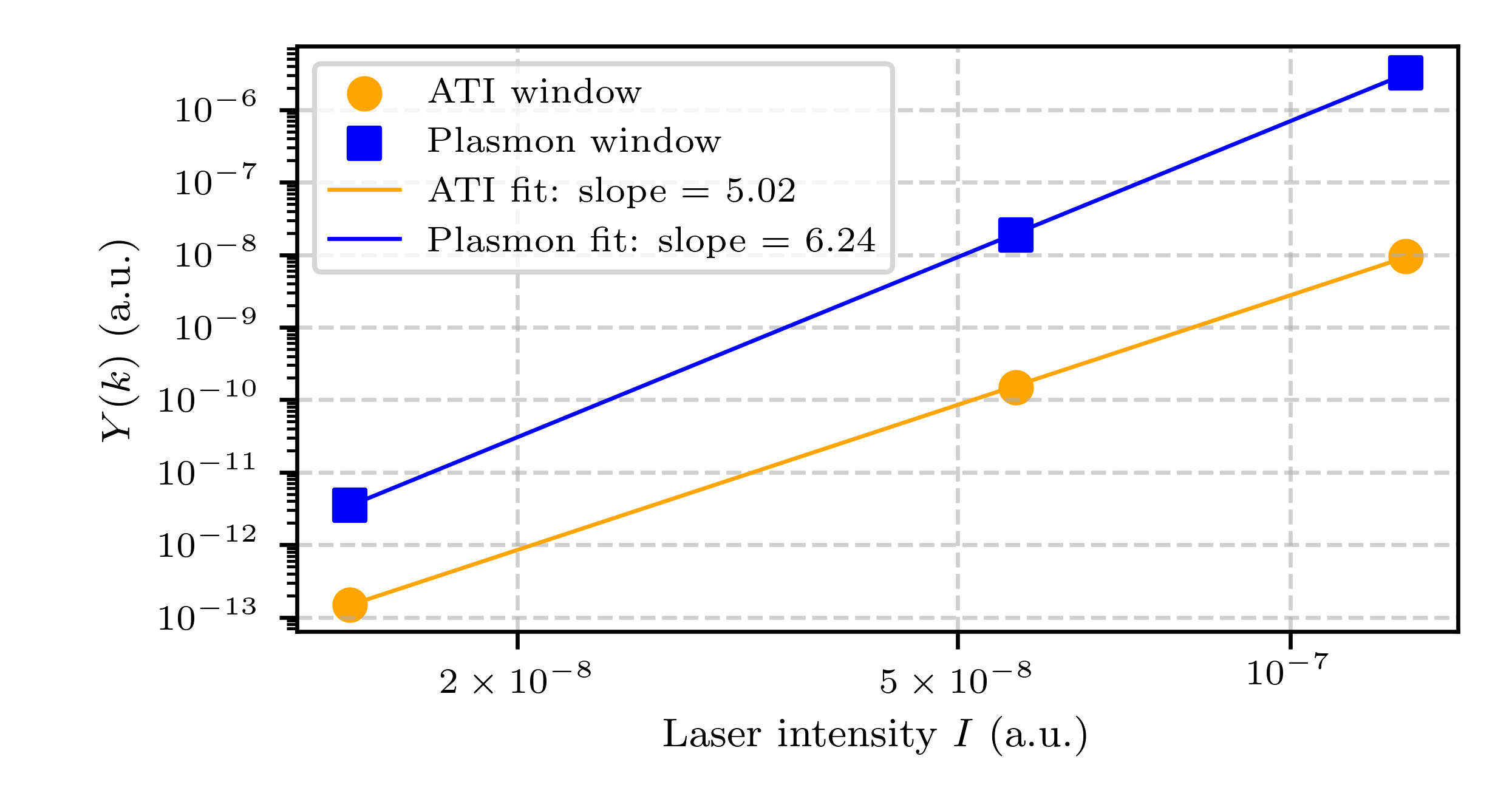}
  \caption{
    \label{fig:int_comp_035}
    Peak yields {\em vs} intensity  for Fig.~\ref{fig:tsurff_comp_035}. The ATI peak scales with $n=5.02$ while the plasmon peak scales with $n=6.24$.
    }
\end{figure}

We show another example with the $\omega_A$-plasmon involved in Figs.~\ref{fig:tsurff_comp_035} and \ref{fig:int_comp_035}. Again, three photons with $\omega_L=0.035$ are required to excite the plasmon. We look at the ATI peak at $\varepsilon_{20}+5\omega_L=0.0041$ and find the expected $n=5$-scaling. For the plasmon peak at $\varepsilon_{20}+2\omega_A=0.041$ we find numerically $n=6.24\simeq 6$, which also fits with the above interpretation that the excitation of the plasmon scales with $I^3$ and the emission process requires another $\omega_A=3\omega_L$, which compounds to an $I^6$-scaling.

\section{Discussion and outlook}
While plasmon-assisted electron emission exists, it remains uncertain whether TDDFT with currently practicable exchange-correlation functionals captures this phenomenon accurately. There are known cases where standard Kohn-Sham TDDFT fails, and more advanced approaches are required \cite{Ruggi2009, Dar2024, Brics2013}. Our results indicate that plasmon-assisted electron emission is likely overemphasized in the present simulations: the associated oscillations should decay more rapidly after the laser pulse. This behavior can be attributed to several limitations of our model. First, Landau damping is suppressed in one dimension. In higher dimensions, the larger density of states would enhance dephasing, leading to a shorter plasmon lifetime. Second, real dissipation channels, such as coupling to phonons or the environment, are absent from our description. Third, the exact exchange-correlation potential, if it were available, would likely accelerate the decay of the plasma oscillations. Plasmon-assisted electron emission may therefore serve as a stringent test case for the development of improved exchange-correlation functionals. Furthermore, it remains unclear whether the scaling of the plasmonic peaks in the photoelectron spectra is captured quantitatively by our TDDFT simulations. There are other examples in strong-field physics where TDDFT predicts qualitatively correct but quantitatively inaccurate features, such as the second plateau in high-harmonic generation \cite{Brics2016}. Future work will thus focus on models amenable to higher-level many-body techniques, allowing for systematic benchmarking of TDDFT in this challenging regime.

\begin{acknowledgments}
We acknowledge funding from the Deutsche Forschungsgemeinschaft via SFB 1477 'Light-Matter Interactions at Interfaces' (project no. 441234705).
\end{acknowledgments}

\bibliography{bib}

\end{document}